\documentclass[french]{gretsi}
\usepackage[utf8]{inputenc}
\usepackage[T1]{fontenc}
\usepackage{dsfont}
\usepackage{amsmath, amssymb, amsfonts}
\usepackage[french]{babel}   
\usepackage{svg}
\usepackage{icomma}
\usepackage{url}

\input alphabet.tex

\usepackage{colortbl}

\def\remHC#1{\textcolor{blue}{\raisebox{-2pt}{$\bullet$}\kern-4pt\footnotemark}\footnotetext{\textcolor{blue}{#1}~}}
\usepackage{soul}

\def\remAM#1{\textcolor{red}{\raisebox{-2pt}{$\bullet$}\kern-4pt\footnotemark}\footnotetext{\textcolor{red}{#1}~}}

\definecolor{darkgreen}{rgb}{0.,0.5,0.}

\newenvironment{itemize*}
    {\begin{itemize}%
      \setlength{\itemsep}{0pt}%
      \setlength{\parskip}{0pt}}%
    {\end{itemize}}

\begin{document}
	
\titre{Réévaluer l'évaluation : classification hyperspectrale sans reconstruction et au-delà des vérités terrain}
\titre{Classification non supervisées d'acquisitions hyperspectrales codées: quelles vérités terrain?}
	
	\auteurs{
		\auteur{Trung-Tin}{Dinh}{tdinh@irap.omp.eu}{1,2}
		\auteur{Hervé}{Carfantan}{herve.carfantan@irap.omp.eu}{1}
		\auteur{Antoine}{Monmayrant}{antoine.monmayrant@laas.fr}{2}
		\auteur{Simon}{Lacroix}{simon.lacroix@laas.fr}{2}
	}
	
	\affils{
		\affil{1}{ Institut de Recherche en Astrophysique et Planétologie (IRAP) -- Université de Toulouse/CNRS/CNES,\\
			14 Avenue Edouard Belin,
			31400 Toulouse, France
		}
		\affil{2}{LAAS-CNRS, Universit\'e de Toulouse, CNRS\\
			7 Avenue du Colonel Roche, 31077 Toulouse, France
		}
	}
	
	\resume{Nous proposons une méthode de classification non supervisée exploitant un nombre réduit d'acquisitions codées issues d'un imageur hyperspectral de type DD-CASSI. 
     Partant d'une modélisation simple de la variabilité spectrale intra-classe, cette approche permet d'identifier des classes et d'estimer des spectres de référence malgré une compression par un facteur dix des données.
     Nous mettons en évidence ici les limites des vérités terrain couramment utilisées pour évaluer ce type de méthode : absence de définition claire de la notion de classe, forte variabilité intra-classe, voire erreurs de classification.
     À partir de la scène \textit{Pavia University}, nous montrons qu'avec des hypothèses simples, il est possible de détecter des régions spectralement plus cohérentes, ce qui souligne la nécessité de repenser l'évaluation des méthodes de classification, notamment dans les scénarios non supervisés.}

	\abstract{We propose an unsupervised classification method using a limited number of coded acquisitions from a DD-CASSI hyperspectral imager. 
     Based on a simple model of intra-class spectral variability, this approach allow to identify classes and estimate reference spectra, despite data compression by a factor of ten.
     Here, we highlight the limitations of the ground truths commonly used to evaluate this type of method: lack of a clear definition of the notion of class, high intra-class variability, and even classification errors.
     Using the \textit{Pavia University} scene, we show that with simple assumptions, it is possible to detect regions that are spectrally more coherent, highlighting the need to rethink the evaluation of classification methods, particularly in unsupervised scenarios.}

	\maketitle


\section{Contexte}
\label{sec:contexte}
Contrairement à l'imagerie classique en couleur (RVB) qui utilise seulement trois bandes spectrales, l'imagerie hyperspectrale enregistre des centaines de bandes étroites couvrant une large gamme de longueurs d'onde. 
Les systèmes d'acquisition traditionnels, tels que les imageurs \textit{pushbroom}, balayent progressivement la scène pour acquérir le cube hyperspectral, ce qui peut nécessiter un temps d'acquisition important et introduire des artefacts dus aux mouvements de la plateforme. 
A l'opposé, des imageurs dits \textit{snapshot} permettent une acquisition instantanée de la scène, réduisant ainsi le temps d'acquisition, en sacrifiant cependant la résolution spatiale. 
En alternative, des imageurs à masques codés permettent d'effectuer des acquisitions en codant des mélanges spectro-spatiaux, sans sacrifier la résolution spatiale. 
En particulier le DD-CASSI \textit{(Double Disperser - Coded Aperture Spectral Snapshot Imager)}~\cite{gehm_single-shot_2007}, préserve la localisation des informations spectrales. 
Il est alors possible d'analyser la scène hyperspectrale à partir de quelques acquisitions, pour des masques différents.
En effet, de nombreux travaux ont proposé des méthodes de reconstruction du cube hyperspectral à partir de telles acquisitions.

Dans ce cadre, nous avons proposé un algorithme de classification spectrale non supervisée~\cite{dinh_statistical_2024}, c'est-à-dire identifiant automatiquement les régions correspondant à des spectres similaires inconnus \textit{a priori}, à partir de quelques acquisitions codées, en s'affranchissant d'une étape de reconstruction.

Cependant, l'évaluation des résultats d'une telle approche par comparaison à une \og la vérité terrain\fg soulève des difficultés, en particulier à cause d'une grande variabilité spectrale intra-classe. 
Cette variabilité peut avoir des origines diverses. 
Le cas le plus simple, que nous avons considéré ici, prend en compte des variations d'intensité globale du spectre d'un même matériau, variations dues à un éclairement variable.
Mais cette variabilité peut bien sûr également provenir de différences spatiales au sein d'un matériau, voire de mélange entre matériaux, ce qui peut poser des questions sur la définition d'une classe.
Enfin, une partie de cette variabilité peut être purement induite par des erreurs de labellisation.

Afin d'illustrer ces problématiques, nous nous appuyons dans ce travail sur une scène hyperspectrale largement utilisée dans la communauté appelée \textit{Pavia University} (Pavia-U)~\cite{gamba_pavia_2001}, mais des observations similaires ont été obtenues sur d'autres jeux de données comme \textit{Indian Pines}~\cite{gamba_pavia_2001}, \textit{CAMCATT} \cite{roupioz_multi-source_2023}.

Dans la section~\ref{sec:modélisation} nous présenterons les modèles, hypothèses et outils exploités pour la méthode de classification non supervisée d'images hyperspectrales à partir de données codées que nous avons proposé~\cite{dinh_statistical_2024}. 
Cela nous amènera à préciser ce que nous définissons comme classe dans une telle méthode.
Dans la section~\ref{sec:pavia} nous analyserons les données de la classifications de référence fournie avec les données hyperspectrales de la scène \textit{Pavia-U}, en particulier en terme de variabilité spectrale.
Enfin, dans la section~\ref{sec:analyse}, nous analyserons sur cette base les résultats obtenus par notre méthode sur cette scène, avec 10 fois moins de données codées que le cube hyperspectral.

\section{Modélisation et outils exploités}

\label{sec:modélisation}

Dans un dispositif d'acquisition codé de type DD-CASSI la scène observée est dispersée spectralement puis projetée à travers un masque codé avant de subir une dispersion inverse et d'être captée par un détecteur.
Pour un objet hyperspectral $\ob$ de dimension $R\times C\times W$ (où $W$ est le nombre de longueurs d'onde) on peut effectuer $A$ acquisitions de dimension $R\times C$ pour $A$ configurations différentes du masque.
L'objectif est d'analyser la scène avec $A\ll W,$ soit beaucoup moins d'acquisitions que de longueurs d'onde du cube (typiquement $A<W/10$).

Afin de combler le manque d'information, il est nécessaire de prendre en compte la corrélation spectro-spatiale de l'objet.
Dans un cadre de classification, on peut considérer que les pixels d'une même région $k$ partagent un spectre de référence~$\sb_k$ avec une éventuelle variabilité.
Dans nos travaux, nous considérons un modèle simple de variabilité spectrale à savoir une variabilité locale d'intensité~\cite{drumetz_blind_2016,hemsley_fast_2022,dinh_statistical_2024}, soit pour $\ob_n\in\eR^W$ le spectre pour le $n$ième pixel de la région $\Rc_k$ : $\ob_n = \psi_n\sb_k,$ où $\psi_n\in\eR$ le coefficient de variation de l'intensité, est supposé proche de 1.
Ce coefficient $\psi_n$ peut aisément s'estimer à partir de l'image panchromatique, elle même directement accessible par le dispositif (acquisition sans masque).

Partant de ce modèle, nous avons proposé une méthode de classification non supervisée~\cite{dinh_statistical_2024}, à partir d'un faible nombre d'acquisitions codées, qui exploite plusieurs ingrédients :
\vspace{-.2cm}

\begin{itemize*}
\setlength\parskip{0pt}
\setlength\itemsep{0pt}
\setlength\topsep{0pt}
\item estimation du spectre de référence $\widehat \sb_k$ par reconstruction sur une région homogène par la méthode SA~\cite{hemsley_fast_2022};
\item estimation du coefficient de variabilité $\widehat\psi_n$ pour les pixels de la région $\Rc_k$;
\item suppression des pixels de trop forte variabilité, pour un seuil $T$: $|1-\widehat\psi_n|<T$;
\item test de Gaussianité sur les résidus $\widehat \db_n -\db_n,$ où $\db_n\in\eR^A,$  pour l'ensemble des pixels $n\in\Rc_k$ d'une région $\Rc_k$ correspond aux données acquises pour le pixel $n$ et $\widehat \db_n$ la prédiction des données pour ce pixel pour le spectre $\widehat\sb_k$ et l'intensité $\widehat\psi_n.$
\end{itemize*}\vspace{-.2cm}

\noindent Avec ces ingrédients, l'algorithme proposé fonctionne en plusieurs étapes :
(1)~Détection de régions homogènes~$\Rc_k$ de~$P\times P$ pixels (avec $P^2>W/A$) et du spectre de référence $\widehat \sb_k$ associé validant le test de Gaussianité et le seuil de faible variabilité ;
(2)~Croissance de la région $\Rc_k$ satisfaisant les mêmes conditions ;
(3)~Fusion de régions satisfaisant les mêmes conditions.
Hormis le nombre de pixels $P^2$ des régions homogènes utilisées pour reconstruire le spectre de référence, seul le seuil de variabilité~$T$ doit être réglé par l'utilisateur.

L'objectif ici n'est pas de présenter en détail cet algorithme mais de mettre en avant l'analyse des résultats obtenus par un algorithme de ce type sur des données pour lesquelles on dispose de \og vérités terrain.\fg

\section{\textit{Pavia-U} et variabilité spectrale}
\label{sec:pavia}

La scène \textit{Pavia-U} pour laquelle quelques régions ont été labellisées et sont utilisées comme \og vérité terrain,\fg a été largement utilisée pour évaluer les performances des algorithmes de classification hyperspectrale, qu'il s'agisse de méthodes supervisées ou non supervisées~\cite{fauvel_advances_2013}, ou de techniques d'apprentissage automatique~\cite{ahmad_hyperspectral_2021}.

\begin{figure}
     \centering     
     \includegraphics[width = \columnwidth]{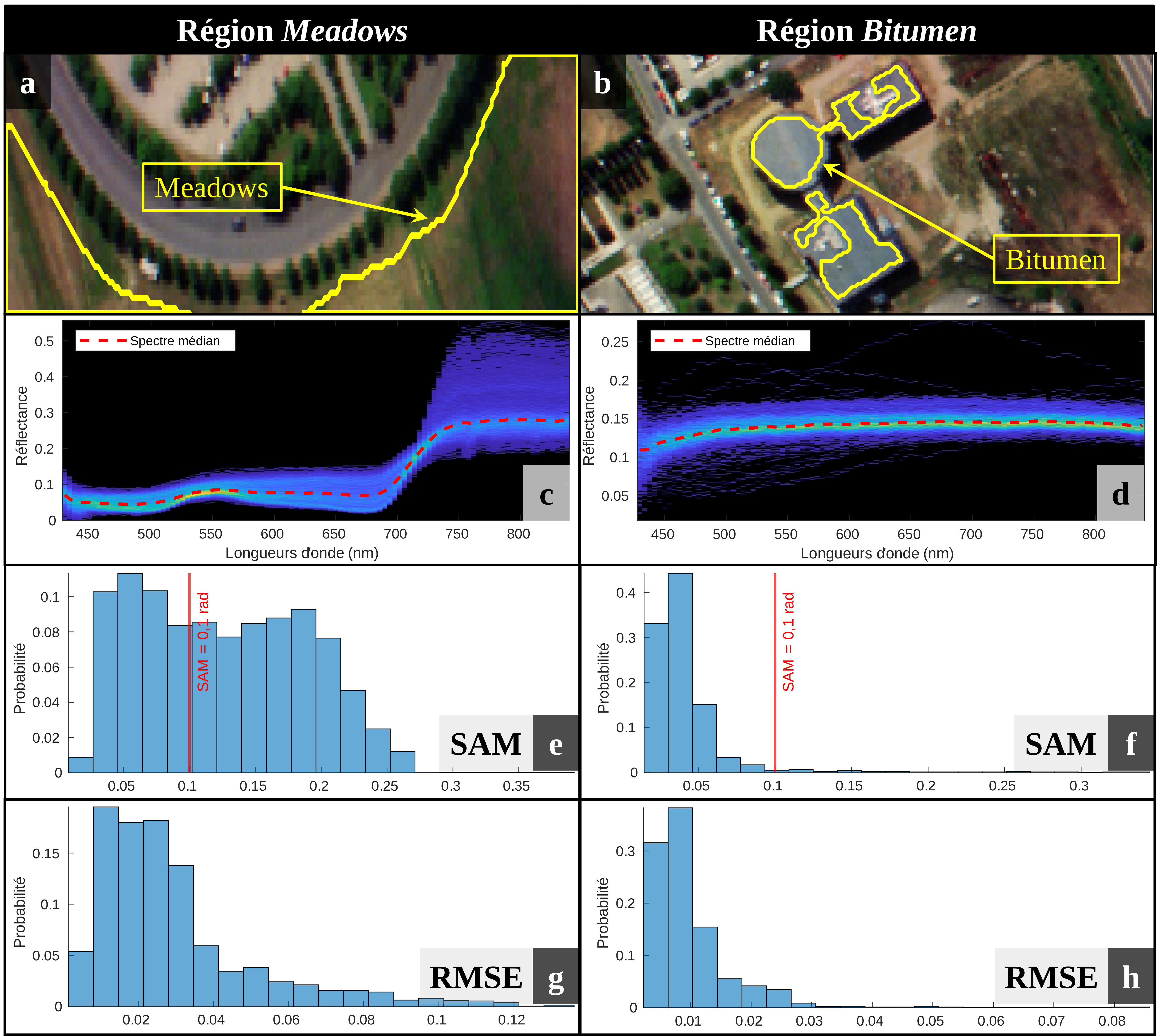}     
     \caption{Spectres extraits de la scène \textit{Pavia-U}. De haut en bas : scène observée en couleurs RVB avec région d'intérêt -- correspondant à une classe de référence -- entourée en jaune \textbf{(a, b)}, l'ensemble des spectres avec les valeurs médianes \textbf{(c, d)}, histogrammes de SAM \textbf{(e, f)} et RMSE \textbf{(g, h)} entre les spectres de chaque région et leurs spectres médians.}
     \label{fig:spectres_paviau}
\end{figure}

Cependant, la variabilité spectrale intra-classe présente dans ces régions labellisées complique la définition même de classes et l'évaluation des résultats de classification.
Pour illustrer cette variabilité, nous extrayons deux petites régions appartenant respectivement aux classes de référence \textit{Meadows} et \textit{Bitumen} fournies avec le jeu de données (voir Fig.~\ref{fig:spectres_paviau}).
En observant l'image en couleurs RVB de la région \textit{Meadows} (Fig.~\ref{fig:spectres_paviau}a), nous constatons une grande diversité de couleurs parmi les pixels, suggérant une variabilité spectrale significative.
Pour la région \textit{Bitumen} (Fig.~\ref{fig:spectres_paviau}b), les pixels apparaissent plus homogènes en couleurs, mais des variations sont également visibles, notamment dans la partie circulaire du bâtiment.

Afin d'analyser l'impact des limites de la vérité terrain sur l'évaluation des résultats, nous utilisons deux métriques couramment employées en imagerie hyperspectrale, appliquées aux spectres de chaque classe en comparaison avec un spectre de référence (choisi ici comme le spectre médian).
La similarité spectrale est quantifiée à l'aide de l'angle spectral (SAM - \textit{Spectral Angle Mapper}), qui est indépendant de l'intensité des spectres.
Cet angle est compris entre $0$ et $\pi$ rad, où 0 indique une similarité parfaite.
Par ailleurs, l'erreur quadratique moyenne (RMSE - \textit{Root Mean Square Error}) mesure l'écart global entre deux spectres. 
Pour des spectres de réflectance, les intensités spectrales sont comprises entre 0 et 1, la valeur du RMSE est également bornée entre 0 et 1, où 0 traduit une parfaite correspondance entre les spectres.

Avec l'analyse des spectres (Fig.~\ref{fig:spectres_paviau}c et \ref{fig:spectres_paviau}d), on constate une importante variation de ces spectres au sein de chaque région, aussi bien en termes de forme que d'intensité, même dans la région \textit{Bitumen} qui paraissait plus homogène dans l'image RVB.
Dans la région \textit{Meadows}, on distingue visuellement au moins trois groupes de spectres de formes différentes. 
Pour \textit{Bitumen}, bien que la majorité des spectres soient proches, certains apparaissent clairement aberrants. 
Cette variabilité se retrouve dans les valeurs de SAM (Fig.~\ref{fig:spectres_paviau}e, \ref{fig:spectres_paviau}f) et de RMSE (Fig.~\ref{fig:spectres_paviau}g, \ref{fig:spectres_paviau}h).
Par exemple, $59,8\%$ des pixels de \textit{Meadows} ont un SAM $> 0,1$ rad, ce qui dépasse le seuil usuel admis pour une même classe dans l'outil de classification d'ENVI \cite{l3harris_geospatial_envi_nodate}. 
Pour \textit{Bitumen}, ce pourcentage est de $2,3\%$, confirmant son homogénéité relative. 
Les RMSE suivent une tendance similaire, traduisant une variabilité plus marquée à \textit{Meadows} qu'à \textit{Bitumen}.

Nous sommes ainsi face à deux situations contrastées : 
la région \textit{Meadows} présente une forte variabilité spectrale, suggérant que la vérité terrain gagnerait à être subdivisée en sous-régions plus homogènes ; à l'inverse, la région Bitumen apparait globalement bien labellisée, hormis quelques spectres aberrants qui pourraient être exclus pour améliorer la cohérence spectrale.

\section{Analyse des résultats}
\label{sec:analyse}

Afin de générer des données codées réalistes à partir de cubes hyperspectraux de référence, nous utilisons le simulateur SIMCA \cite{rouxel_accurate_2024}.
Ce simulateur intègre les paramètres instrumentaux du système d'acquisition, en prenant en compte les distorsions optiques ainsi que les caractéristiques du masque codé.
Les données codées ont été simulées sur $A = 10$ acquisitions, soit une quantité de données dix fois inférieure à celle du cube hyperspectral complet ($W = 103$).

\begin{figure}
     \centering      
     \includegraphics[width = \columnwidth]{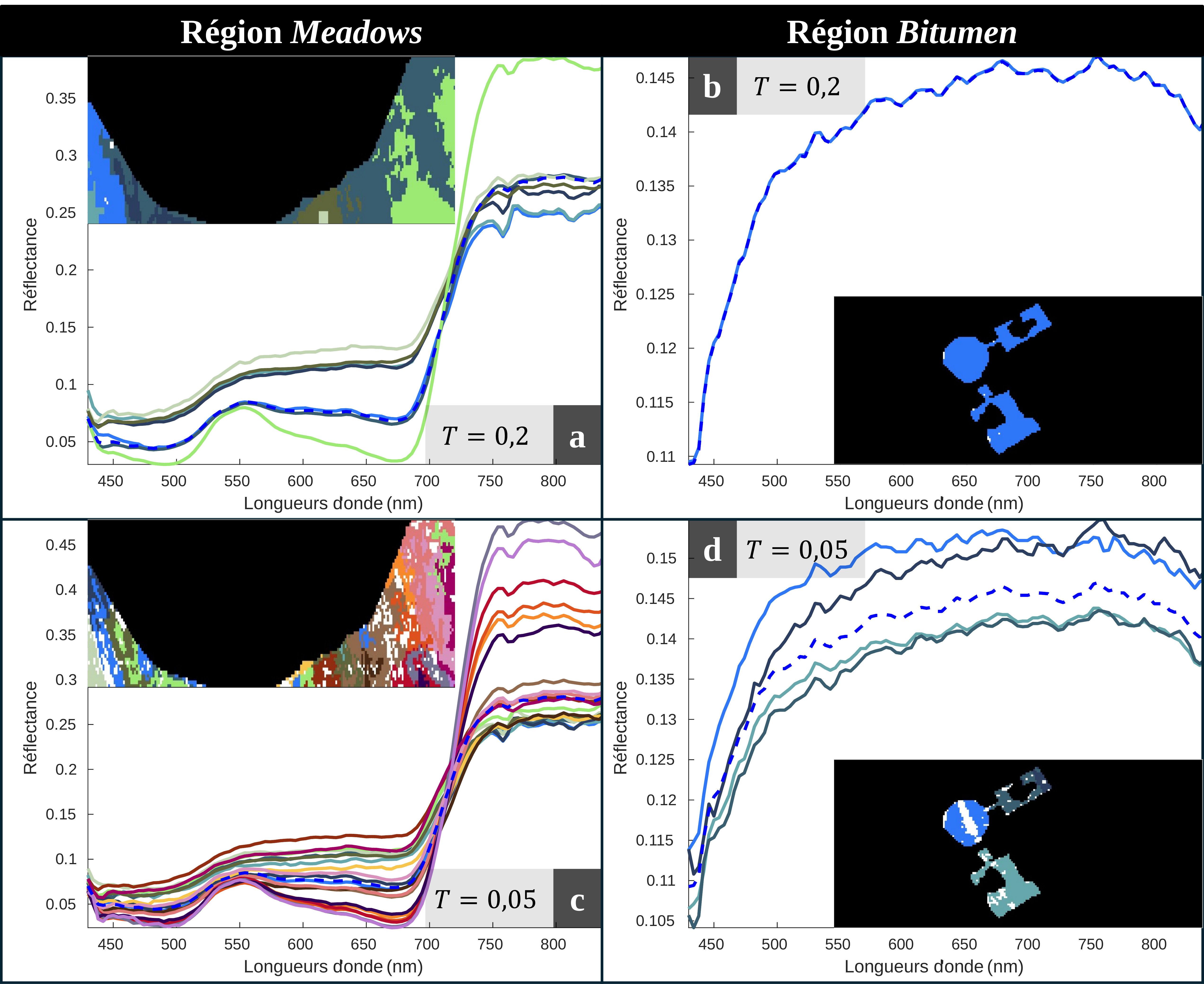}     
     \caption{Résultats de classification (régions classées et spectres médians correspondants) pour différentes valeurs de $T.$ Images des régions labellisées: \textit{couleurs} - régions homogènes, \textit{blanc} - régions non classées, \textit{noir} - non concernées; spectres médians: \textit{trait pointillé} - spectre de référence, \textit{trait plein} - spectre médian de chaque région homogène.     
     }
     \label{fig:label_classif_s_ref}
\end{figure}

Notre algorithme de classification non supervisée est ensuite appliqué aux données codées générées, en fixant une valeur de~$T$. 
Pour chaque valeur de~$T$, nous obtenons un ensemble de régions homogènes et des régions non classées, ces dernières correspondant aux zones de mélange ou à des zones où l'information est insuffisante.
Dans la Fig.~\ref{fig:label_classif_s_ref}, les régions homogènes sont représentées en couleurs, les pixels blancs indiquent des pixels non classés, et les pixels noirs, en dehors de la région d'intérêt, sont exclus de la comparaison.

Outre les régions labellisées, notre algorithme fournit un spectre reconstruit régularisé pour chaque région homogène, basé sur la méthode SA \cite{hemsley_fast_2022}. 
Toutefois, dans le contexte de classification, nous choisissons d'utiliser, pour chaque région détectée, le spectre médian calculé à partir des spectres complets des pixels de la région. 
Ce choix permet une comparaison équitable avec la classe de référence, elle-même représentée par son spectre médian. 
En effet, le spectre régularisé n'est disponible que pour les régions issues de notre algorithme, et ne peut être estimé pour les classes de la vérité terrain, ce qui introduirait un biais dans l'évaluation. 
Les spectres médians ainsi obtenus sont tracés dans la Fig.~\ref{fig:label_classif_s_ref} avec les mêmes couleurs que les régions correspondantes, et le spectre de la vérité terrain est représenté en trait pointillé.

\begin{figure}[ht]
     \centering     
     \includegraphics[width = \columnwidth]{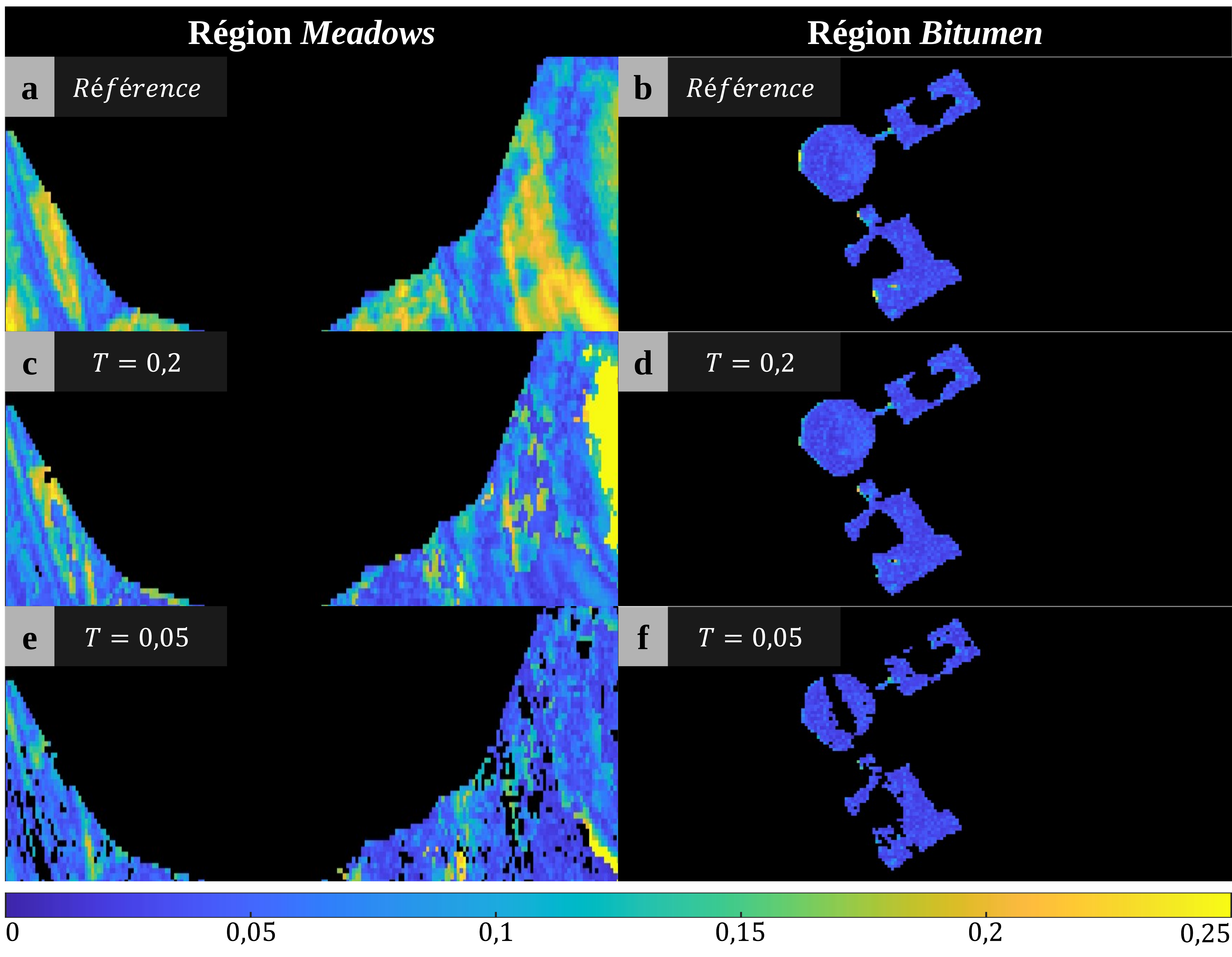}
     \caption{Cartes de SAM de la vérité terrain pour différentes valeurs de $T.$}
     \label{fig:carte_sam} 
\end{figure}

Pour la région \textit{Meadows}, notre algorithme a identifié 7 régions distinctes avec $T = 0,2$ et 20 régions pour $T = 0,05$ (Fig.~\ref{fig:label_classif_s_ref}a et \ref{fig:label_classif_s_ref}c). 
Leurs spectres médians diffèrent notablement de la référence, mais sont cohérents avec notre observation de la variabilité spectrale et de l'existence de plusieurs spectres différents dans la Fig.~\ref{fig:spectres_paviau}c.
Pour \textit{Bitumen}, notre algorithme a identifié 1 région homogène avec $T = 0,2$ et 4 régions avec $T = 0,05$ (Fig.~\ref{fig:label_classif_s_ref}b et \ref{fig:label_classif_s_ref}d). 
Cela est peu surprenant, car cette région est plus homogène que celle de \textit{Meadows}.
Pour $T = 0,2$, quelques pixels aberrants ont été supprimés (Fig~\ref{fig:carte_sam}d),
mais la majorité des pixels sont classés dans une même région homogène correspondant dans les grandes lignes à la classification de référence.
Notons que plus le seuil $T$ est petit, plus le nombre de régions homogènes détectées est important, et plus les spectres médians de ces régions sont différents du spectre de référence de la classe de référence.

Pour quantifier la variabilité intra-classe, dans chaque région labellisée, nous calculons le SAM entre les spectres des pixels et le spectre médian. 
Les cartes de SAM correspondantes sont présentées en Fig.~\ref{fig:carte_sam}.
Pour la région \textit{Meadows}, la carte SAM associée à la classification de référence (Fig.~\ref{fig:carte_sam}a) montre une forte variabilité spectrale, avec de larges zones présentant des valeurs élevées, certaines étant même nettement discernables sur l'image RVB par des couleurs distinctes. 
En appliquant notre méthode avec $T = 0,2$ (Fig.~\ref{fig:carte_sam}c), on observe une nette diminution des valeurs de SAM dans la majorité des régions, traduisant une plus faible variabilité intra-classe. 
Pour $T = 0,05$ (Fig.~\ref{fig:carte_sam}e), la réduction est encore plus marquée, sauf dans une petite zone localisée où il existe probablement un mélange complexe de matériaux.
Pour la région \textit{Bitumen}, la carte SAM de la classe de référence (Fig.~\ref{fig:carte_sam}b) montre une variabilité globalement faible, mais quelques petites zones isolées présentent des valeurs plus élevées. 
Avec $T = 0,2$ (Fig.~\ref{fig:carte_sam}e), notre méthode élimine ces pixels aberrants, réduisant la variabilité.
Enfin, pour T = 0,05 (Fig.~\ref{fig:carte_sam}f), la carte de SAM devient encore plus homogène. 
Ainsi, l'analyse des cartes de SAM montre que notre méthode permet de mieux détecter les régions véritablement homogènes, présentant une faible variabilité intra-classe, contrairement à la classification de référence qui regroupe parfois des régions présentant des spectres dissemblables, voire des spectres aberrants.

\begin{figure}[ht]
     \centering     
     \includegraphics[width = \columnwidth]{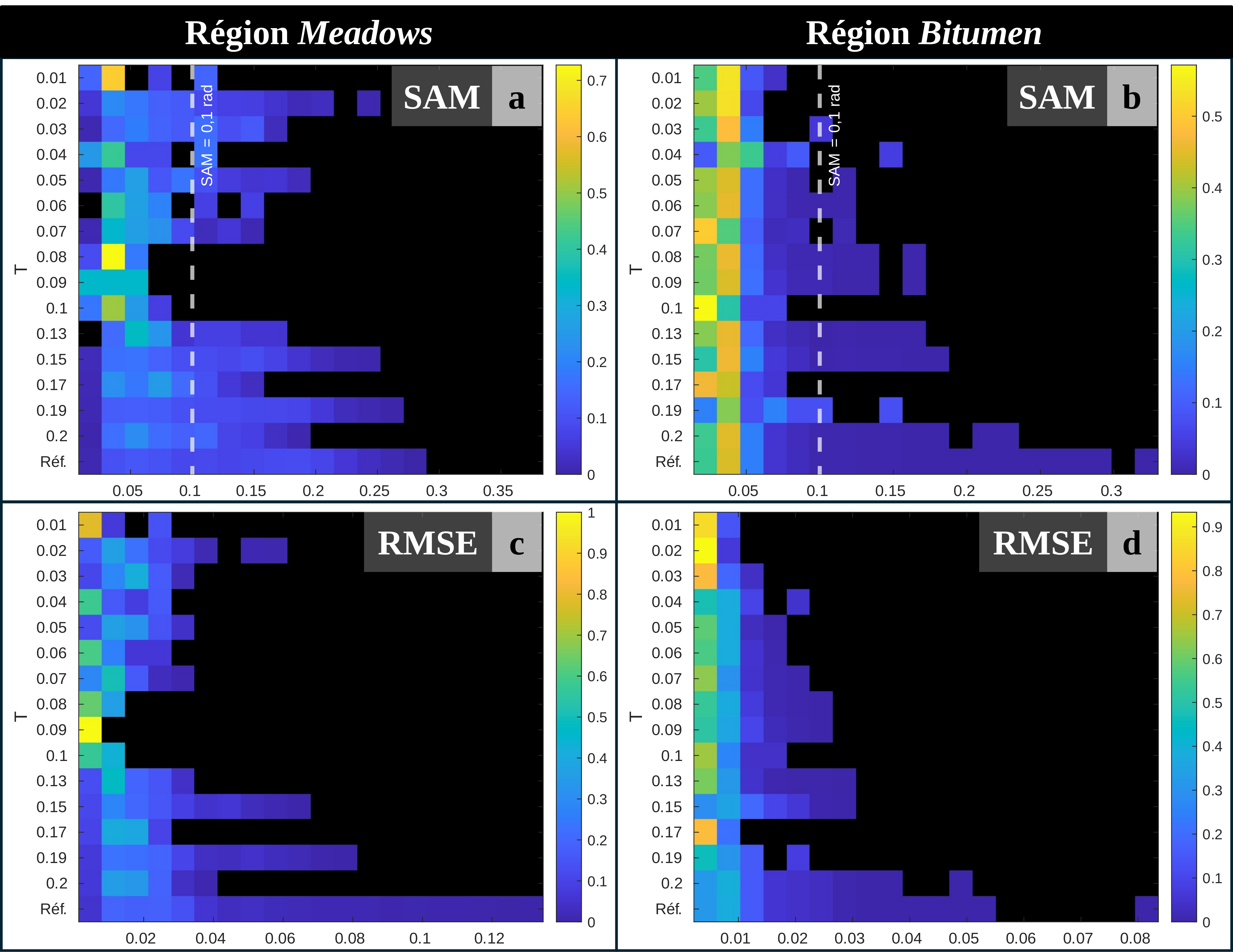}
     \caption{Histogrammes de probabilité des SAM et RMSE pour différentes valeurs de $T$ et de la vérité terrain.}
     \label{fig:metriques_o_paviau}
\end{figure}

Les histogrammes de probabilité (Fig.~\ref{fig:metriques_o_paviau}) permettent de visualiser la distribution des valeurs de SAM et de RMSE suivant la valeur du seuil~$T$, ainsi que pour la classification de référence.
Les histogrammes sont présentés sous forme d'images où chaque ligne correspond à une valeur de $T$, la dernière ligne représentant les valeurs issues de la classification de référence ; les valeurs des probabilités étant indiquées par des couleurs.
Nous constatons que les distributions de SAM et de RMSE des régions classées par notre algorithme sont plus concentrées autour de 0 et présentent des valeurs moyennes plus faibles que celles des régions de la classification de référence, traduisant une meilleure cohérence spectrale intra-classe.

Pour la région \textit{Meadows}, on observe que les histogrammes issus de notre classification deviennent progressivement plus étroits et concentrés autour de $0$ lorsque $T$ diminue jusqu'à un seuil $T \sim 0,09$, ce qui reflète une diminution de la variabilité spectrale intra-classe. 
En deçà de ce seuil, les distributions s'élargissent à nouveau.
Cela est dû à des difficultés rencontrées par notre algorithme pour l'estimation de spectres de référence sur certaines régions de petite taille, ce qui déborde du cadre de cette étude. 
Néanmoins, l'histogramme associé à la classification de référence reste globalement plus large, confirmant sa moindre homogénéité spectrale.

Pour la région \textit{Bitumen}, bien que la vérité terrain montre déjà une certaine homogénéité (distributions initialement resserrées), nos résultats en réduisant la valeur de $T$ permettent de réduire encore davantage la dispersion des valeurs de SAM et RMSE (Fig.~\ref{fig:metriques_o_paviau}b et \ref{fig:metriques_o_paviau}d) en excluant progressivement des spectres aberrants. 

Ces résultats confirment que notre méthode identifie des régions plus cohérentes du point de vue de variabilité spectrale, tandis que la vérité terrain, utilisée comme référence, ne reflète pas toujours fidèlement l'homogénéité spectrale réelle des régions, regroupant parfois des spectres hétérogènes, ce qui peut fausser l'évaluation.

\section{Conclusions}
\label{sec:discussion}
Ce travail met en évidence les limites des vérités terrain dans les jeux de données hyperspectrales, souvent considérées comme des références absolues, notamment pour l'évaluation des méthodes de classification non supervisée.
Tout en cherchant à analyser les résultats d'un algorithme spécifique de classification à partir d'acquisitions codées, nous avons montré que ces vérités terrain, bien qu'issues d'annotations expertes, peuvent introduire un biais significatif dans l'évaluation, en regroupant sous une même classe des pixels aux signatures spectrales pourtant dissemblables.
Ce problème n'est pas spécifique à une scène donnée : il se retrouve aussi bien dans les bases de données classiques~\cite{gamba_pavia_2001} \textit{(Indian Pines, Pavia Center, Salinas, \dots)} que dans des jeux récents et rigoureusement construits comme \textit{CAMCATT} \cite{roupioz_multi-source_2023}, où des spectres de référence sont pourtant mesurés \textit{in situ} à l'aide de spectromètres.

Plus généralement, la modélisation et la prise en compte de variabilité spectrale intra-classe est un enjeu difficile, qui remet en question la définition de la notion de classe et donc celle de vérité terrain.
Comme nous l'avons illustré, cela rend délicat l'évaluation des méthodes de segmentation, mais on peut également se questionner sur l'influence de telles définitions de classes sur des algorithmes d'apprentissage s'appuyant sur ces \og vérités terrain.\fg

\footnotesize
\bibliographystyle{gretsi}
\bibliography{DINH_Gretsi_v2}

\end{document}